\documentclass[11pt]{article}
\usepackage[english]{babel}
\usepackage[dvips]{graphics,color}
\usepackage{newlfont,epsfig}
\usepackage{graphicx}
\title{Atom-field transfer of coherence in a two-photon micromaser assisted by a classical field}

\author{A.F. Gomes$^{a,b}$ and A. Vidiella-Barranco$^b$}

\date{$^a$Funda\c c\~ao Educacional de Barretos, C.P. 16, 14783-226, Barretos SP Brazil\\
$^b$Instituto de F\'\i sica ``Gleb Wataghin'', Universidade Estadual de Campinas, 
13083-970 Campinas SP Brazil}

\begin{document}
\maketitle



\begin{abstract}
We investigate the transfer of coherence from atoms to a cavity field initially in a statistical mixture 
in a two-photon micromaser arrangement. The field is progressively modified from a maximum entropy state
(thermal state) towards an almost pure state (entropy close to zero) due to its interaction with atoms 
sent across the cavity. We trace over the atomic variables, i.e., the atomic states are not collapsed by a detector after they leave the cavity. We find that by applying an external classical driving field it is possible to substantially increase the field purity without the need of previously preparing the atoms in a superposition of 
their energy eigenstates. We also discuss some of the nonclassical features of the resulting field.
\end{abstract}



\section{Introduction}

In the past years there have been important advances regarding the control over both isolated
atoms and electromagnetic fields. This is relevant not only for the investigation of quantum behaviour,
but also because of potential applications. It is of particular importance the generation of 
pure states of the field. With respect to this, high Q single-mode cavities have been shown to be adequate 
systems for the generation and manipulation of quantum states of light \cite{haroche}. In general, atoms conveniently 
prepared are injected inside a cavity, and their coupling to the quantized cavity field makes possible to modify it
in such a way that nonclassical states of the field may be generated \cite{haroche1}. Other nonclassical effects, such 
as atomic dipole squeezing, may also occur in a micromaser system, as discussed in \cite{rao}. 
The cavity itself is normally
cooled down to its vacuum state, and the field build up may be accomplished by coherent addition of photons to it.
Nevertheless, the radiation field in most ordinary situations, e.g., in thermal equilibrium, happens to be in a 
mixed state. It would be therefore interesting to investigate the construction of quantum states of light departing 
from mixed (thermal) states rather than the vacuum state. In previous
papers it has been shown that in micromasers undergoing either one-photon \cite{hector} or two-photon \cite{alvaro1} 
transitions, field states with a modest degree of purity may be generated from highly mixed states, even if one does 
not perform conditional measurements on the atoms after they leave the cavity. Despite of the 
fact that two-photon transitions \cite{alvaro1} are more efficient compared to one-photon ones regarding 
the atom-field transfer of coherence process, the degree of mixedness of the resulting field in those schemes is still 
considerably high. It would therefore be interesting to seek for ways of improving the process of transfer of
coherence from atoms to the cavity field. Here we
consider a situation in which the cavity is driven by an external field, so that 
each incoming atom simultaneously interacts with the cavity (quantized) field as well as with 
the (classical) external driving field. We may find in the literature discussions about several aspects  
of the action of an external field on the system atom-quantized field. In fact its presence 
brings out some nonclassical effects, such as large time scale revivals (super-revivals)
\cite{dutra} and enhancement of quadrature squeezing \cite{gao}, for instance. 
There is also a proposal of a scheme of single-photon states generation \cite{nha} using an external driving 
field. Here we study the influence of an external field on the process of transfer of coherence from atoms to 
the field in a two-photon micromaser. Our aim is the construction of quantum states of light as pure as possible
starting from a mixed state. We find that the driving field brings us several advantages: a simpler 
preparation of the atomic states is required (excited state $|e\rangle$ instead of a coherent superposition 
$|e\rangle + |g\rangle$); the transfer of coherence process is faster than in the case in which the external 
field is not present; the resulting field is very close to a pure state and has nonclassical properties as well. 
We are considering an ideal cavity, although we are aware that cavity losses are an important source of 
decoherence that will for sure degrade the process of transfer of coherence. We are not including losses in our analysis because our main aim is to investigate the main features of the influence of the driving field.
In order to characterize the generated field we employ representations of the field in number (photon number 
distribution) and coherent ($Q$-function) states as well as the second order correlation function, 
$g^{(2)}(0)$. The paper is organized as follows: in section 2 we solve the model and calculate the field density 
matrix after after the passage of $N$ atoms through 
the cavity; in section 3 we discuss our numerical results concerning the degree of field purification using the linear
entropy (mixedness parameter). We also analize the field nonclassical properties; in section 4 we present our 
conclusions.

\begin{figure}
\vspace{0.5cm}
\begin{center}
{\includegraphics[height=5.5truecm,width=7.5truecm]{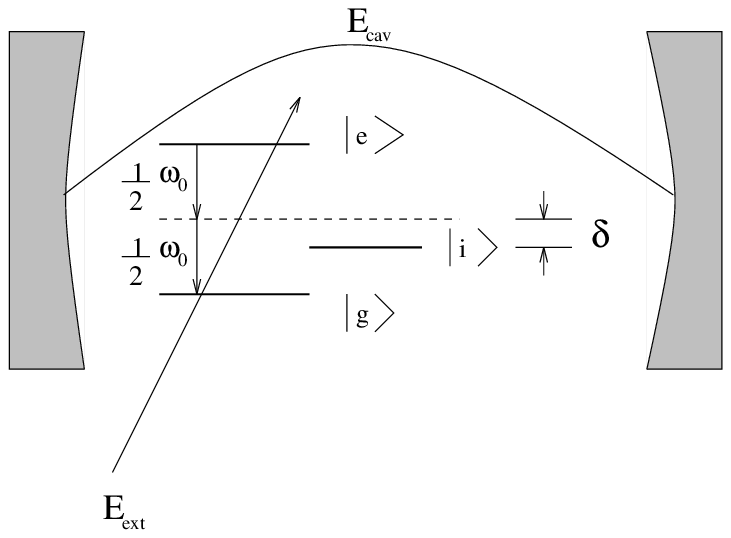}}
\end{center}
\vspace{0.1cm}
\caption{Schematic representation of the two-photon micromaser setup.}
\label{fig1}
\end{figure}

\section{The model}

We consider a two-photon micromaser in which the injected (three-level) atoms are coupled to the cavity field as 
well as to a classical driving field, in a configuration shown in figure 1. 
If the intermediate atomic level $|i\rangle$ is highly detuned from the field, i.e., if $\delta=E_i - (E_e + E_g)/2$ 
is large enough, it may be adiabatically eliminated so that we obtain the following effective hamiltonian
\begin{eqnarray}
\hat{H}&=&\hbar \omega \hat a^{\dagger} \hat a + \hbar \left[\frac{\omega_0} {2}+\chi (\hat a^{\dagger} 
+ \varepsilon e^{-i \omega' t})(\hat a + \varepsilon e^{i \omega' t} ) \right]\sigma_z\nonumber\nonumber\\
&+& \lambda\left[(\hat a^{\dagger} + \varepsilon e^{-i \omega' t} )^2 \sigma_{-}
+ (\hat a + \varepsilon e^{i \omega' t})^2 \sigma_{+} \right],\label{hamil1}
\end{eqnarray}
where $\lambda=2 g^2/\delta$ is the effective coupling constant
(being $g=g_{ig}=g_{ei}$ the coupling constants related to one-photon transitions), and 
$\sigma_+ = \sigma_{eg}$, $\sigma_- = \sigma_{ge}$ $\sigma_z=\sigma_{ee}-\sigma_{gg}$ are the atomic transition
operators. The parameter $\chi=2 g^2/\delta$ is the dynamical Stark effect coefficient. The cavity field has
frequency $\omega$, and the classical driving field, of frequency $\omega'$ has an amplitude $\varepsilon$
taken real for simplicity. We assume that there is a detuning $\Delta$ between the cavity field and the atomic 
transition of frequency $\omega_0/2$, i.e., $\Delta= \omega_0 - 2 \omega$, although the classical field itself is 
resonant with the atom, i.e., $\omega'=\omega_0/2$. We may write the evolution operator relative to the hamiltonian in equation 
(\ref{hamil1}) as 
\begin{equation}
\hat U(t) = \hat D^\dagger(\varepsilon^*)\hspace{0.2cm} 
\hat U_{tp}^\dagger(t)\hspace{0.2cm}\hat D(\varepsilon),
\label{DUD}
\end{equation}
\noindent where 
\begin{eqnarray}
\hat U_{tp}(t)=\exp\left[-i\lambda t \left( \frac{\Delta}{2\lambda} \sigma_z + \frac{\chi}
 {\lambda} \hat a^{\dagger} \hat a \sigma_z 
+ \hat a^{\dagger 2} \sigma_{-}
+ \hat a^2 \sigma_{+} \right)\right],
\end{eqnarray}
is the evolution operator for the two-photon Jaynes-Cummings model and 
$\hat D(\varepsilon) = \exp(\varepsilon \hat a^\dagger - \varepsilon^* \hat a)$ is Glauber's displacement
operator.

We assume that the cavity field is initially in a thermal state with mean photon number $\overline{n}$
\begin{equation}
\hat \rho^f(0)=\sum_{n=0}^\infty\frac{\overline{n}^{n}}{(\overline{n}+1)^{n+1}} 
|n\rangle \langle n| \label{infield}
\end{equation} 
and the atom in a superposition of atomic energy eigenstates $|g\rangle$ and $|e\rangle$ states
\begin{equation}
|\phi_a \rangle=b|g\rangle+ae^{i\phi}|e\rangle,\label{atomstate}
\end{equation}
where the coefficients $a$ and $b$ have been taken real for simplicity.
Therefore the atom-field system is initially prepared in the product state 
$\hat{\rho}(0)= \hat \rho^f(0)\otimes |\phi_a \rangle\langle\phi_a |$.

After the atom-field system has evolved during a time $t$, the joint density operator will be 
$\hat{\rho}(t)= \hat U(t)^{\dagger} \hat{\rho}(0)\hat U(t)$. We are interested in keeping the interaction time as short as possible,
in order to minimize the effects of dissipation. In our scheme there are no conditional measurements, 
i.e., the atomic state is not collapsed by a detector and the atoms just fly away after crossing the cavity. 
We are mostly interested in the dynamics of the field, so that we trace over 
the atomic degrees of freedom in order to obtain the field density operator at a time $t$, 
or $\hat{\rho}^f(t)=Tr_a \hat{\rho}(t)$. As in reference 
\cite{alvaro1}, we may calculate the density operator in the number state basis, $\rho_N ^f(n,n')\equiv \langle 
n|\hat{\rho}_N^f(t)|n'\rangle$, after $N$ atoms prepared in superposition states as in equation (\ref{atomstate}) have 
crossed the cavity   
\begin{eqnarray}
\rho_N ^c(n,n')&=&\sum_{m,m'} \rho^c_{N-1}(m,m') \sum_{j,j'} e_{j,m} e^{*}_{j',m'}
\nonumber\\
&[&\hspace{0.3cm}[\hspace{0.1cm}a^2 \alpha_j(\gamma)\alpha_{j'}^{\dagger}(\gamma) 
+b^2\alpha_j(\epsilon)
\alpha_{j'}^{\dagger}(\epsilon)\hspace{0.1cm}]e_{j,n} e^{*}_{n',j'}\nonumber\\
&+&b^2\beta_j(\gamma)\beta_{j'}(\gamma)\sqrt{(j+2)(j+1)(j'+2)(j'+1)} 
\hspace{0.3cm}e_{j+2,n} e^{*}_{n',j+2}\nonumber\\
&+&a^2 \beta_j(\epsilon)\beta_{j'}(\epsilon)\sqrt{j(j-1)j'(j'-1)}
\hspace{0.3cm}e_{j-2,n} e^{*}_{n',j-2}\nonumber\\
&+&iab e^{-i\phi}\alpha_j(\gamma)\beta_{j'}(\gamma)\sqrt{(j'+2)(j'+1)}
\hspace{0.3cm}e_{j,n} e^{*}_{n',j'+2}\nonumber\\
&+&iab e^{i\phi}\alpha_j(\epsilon)\beta_j'(\epsilon)\sqrt{j'(j'-1)}
\hspace{0.3cm}e_{j,n} e^{*}_{n',j'-2}\nonumber\\
&-&iab e^{i\phi}\beta_j(\gamma)\alpha_{j'}^{\dagger}(\gamma)\sqrt{(j+2)(j+1)}
\hspace{0.3cm}e_{j+2,n} e^{*}_{n',j'}\nonumber\\
&-&iab e^{-i\phi}\beta_j(\epsilon)\alpha_{j'}^{\dagger}(\epsilon)
\sqrt{j(j-1)}\hspace{0.3cm}e_{j-2,n} e^{*}_{n',j'}\hspace{0.3cm}],\label{rhocnf}
\end{eqnarray}
\noindent with $e_{j,n}$ and $e^{*}_{n',j'}$ given by
\begin{eqnarray}
 e_{(j,n)} &=& \langle j|\varepsilon;n\rangle = e^{\frac{-|\varepsilon|^2} {2}}\hspace{0.2cm} 
 \varepsilon^{j-n} \hspace{0.2cm}\sqrt{\frac{n!} {j!}}\hspace{0.2cm} \mathcal{L}_{n}^{j-n}
 (|\varepsilon|^2)\nonumber\\\nonumber\\
e^{*}_{(n',j')} &=& \langle n';\varepsilon|j'\rangle = e^{\frac{-|\varepsilon|^2} {2}}\hspace{0.2cm} 
(\varepsilon^{*})^{n'-j'} \hspace{0.2cm}\sqrt{\frac{j'!} {n'!}}\hspace{0.2cm} 
\mathcal{L}_{j'}^{n'-j'}(|\varepsilon|^2), \label{ef}
\end{eqnarray}
where $|\varepsilon;n\rangle = \hat D (\epsilon) |n\rangle$ are displaced number states and 
$\mathcal{L}_{n}^{j-n}$ are the associated Laguerre polynomials. The coefficients
$ {\alpha}_n(\gamma)$ e $ {\beta}_n(\epsilon)$ are
\begin{eqnarray}
 {\alpha}_n(\gamma) &=& \cos(  \gamma_n \lambda t)+i \frac{\sin (  \gamma_n 
\lambda t)} {  \gamma_n} \left(\frac{\Delta}{2 \lambda} + \frac{\chi} {\lambda} (  n + 1) \right), \label{alfak1}
\\\nonumber\\
 {\alpha}_n(\epsilon) &=& \cos(  \epsilon_n \lambda t)+i \frac{\sin (  \epsilon_n 
\lambda t)} {  \epsilon_n} \left(\frac{\Delta}{2 \lambda} + \frac{\chi} {\lambda} (  n - 1) \right), \label{alfak2}
\\\nonumber\\
 {\beta}_n(\gamma) &=& i   a^{\dagger 2} \frac{\sin (  \gamma_n \lambda t)} {  \gamma_n}, 
\label{betak1}
\\\nonumber\\
 {\beta}_n(\epsilon) &=& i   a^2 \frac{\sin (  \epsilon_n \lambda t)} {  \epsilon_n}, 
\label{betak2}
\end{eqnarray}
\noindent with
\begin{eqnarray}
  \gamma_n^2&=&\left(\frac{\Delta}{2 \lambda} + \frac{\chi} {\lambda} (  n + 1)\right)^2
+(  n+1)(  n+2), \label{gamak} \\\nonumber\nonumber\\
  \epsilon_n^2&=&\left(\frac{\Delta}{2 \lambda} + \frac{\chi} {\lambda} (  n - 1)\right)^2
+  n (  n-1). \label{epsilonk}
\end{eqnarray}

A particularly important result is that in order to transfer coherence from atoms to the field, 
there is no need of having the atoms initially prepared in a superposition of their energy 
eigenstates [see equation (\ref{atomstate})]. A simpler atomic preparation, e.g. $|e\rangle$, 
is sufficient, making unnecessary the previous preparation in a Ramsey zone normally employed to generate
superpositions of atomic states, as in \cite{alvaro1}. If the atom is initially prepared in state 
$|e\rangle$ ($b=0$), the resulting density matrix will be
\begin{eqnarray}
\rho_N ^f(n,n')&=&\sum_{m,m'} \rho^f_{N-1}(m,m') \sum_{j,j'} \hspace{0.2cm}e_{j,m} \hspace{0.2cm} e^{*}_{j',m'}
[\hspace{0.25cm}a^2 \alpha_j(\gamma)\alpha_{j'}^{\dagger}(\gamma)
\hspace{0.2cm} e_{j,n} \hspace{0.2cm} e^{*}_{n',j'}\nonumber\\
&+&a^2 \beta_j(\epsilon)\beta_{j'}(\epsilon)\sqrt{j(j-1)j'(j'-1)}
\hspace{0.2cm} e_{j-2,n} \hspace{0.2cm} e^{*}_{n',j-2}\hspace{0.25cm}].\label{rhocnf1}
\end{eqnarray}

We note that its off-diagonal elements may be populated, i.e., $\rho_N ^f(n,n')$ might be nonzero 
(for $n\neq n'$) even for an initial diagonal density matrix as the one in equation (\ref{infield}). Because of 
the presence of the classical field ($\epsilon \neq 0$), there are now sums over $j,j'$ and $m,m'$, which make
possible to have $\rho_N ^f(n,n')\neq 0$. In the present scheme the action of the classical field on the atoms 
is simultaneous with the process of transfer of coherence to the field, differently from previously discussed 
schemes \cite{hector,alvaro1}, which rely upon the previous preparation of the atomic state in a Ramsey zone before
they enter the cavity. We would like to remark that as the field turns less mixed, the atoms,
initially prepared in pure states leave the cavity essentially in mixed states.
We have therefore a method of transfering coherence from atoms to the radiation field assisted by the action of a classical field.

\section{Results}

\subsection{Field state purification}

We now adopt a procedure similar to the one used in reference \cite{alvaro1}. Initially, 
we fix the values of parameters $\Delta$ (detuning) and $\chi$ (Stark shift coefficient).
We are going to look for a resulting  field as pure as possible, having as a condition that 
there is no reduction in the field energy. We consider the situation in which $N$ atoms have already 
crossed the cavity, in such a way that each atom ``gives some coherence'' to the field.
The field purity is normally quantified by the parameter known as linear entropy $\zeta$  
        \[
        \zeta = 1 - Tr[(\rho^f)^2] = 1 - \sum_{n,n'} |\rho_N ^f(n,n')|^2.
\]
For pure states $\zeta=0$, and for statistical mixtures of pure states we have $\zeta > 0$, which means that it 
behaves similarly as the von Neumann entropy $S=-Tr[\rho\, ln (\rho)]$. 
We will numerically determine under which conditions it is possible to minimize $\zeta$. 
The initial field is a thermal state (\ref{infield}) having mean photon number $\overline{n}=5$ and highly mixed
i.e., having a large linear entropy. 
The atoms  entering the cavity are previously prepared in their excited state $|e\rangle$. 
We have calculated the field evolution for a 
range of times, in order to find 
the minimum interaction time which minimizes $\zeta$ (maximizes the field purity)  
after $N$ atoms have crossed the cavity. For simplicity we have considered the same interaction time for each atom. 
The optimum interaction time in this 
case is $T\approx 8.9/\lambda$. In figure 2 we have a plot of the linear entropy $\zeta$ as a function of the total
number of atoms $N$ crossing the cavity, for different values of the classical field amplitude 
$\varepsilon$. 
We note that the linear entropy is very sensitive to $\varepsilon$. For instance, for
$\varepsilon=2.0$, the field purity considerably increases for a relatively small number of atoms. In this case 
$\zeta$ reaches a minimum value $\zeta_{min}\approx 0.18$. If $\varepsilon=1.0$, a considerably purer field is generated, ($\zeta_{min}\approx 0.07$), although it takes longer for $\zeta$ to reach its minimum value 
after $\sim 100$ atoms 
\vspace{0.5cm}
\begin{figure}[h]
\begin{center}
{\includegraphics[height=6truecm,width=4.7truecm]{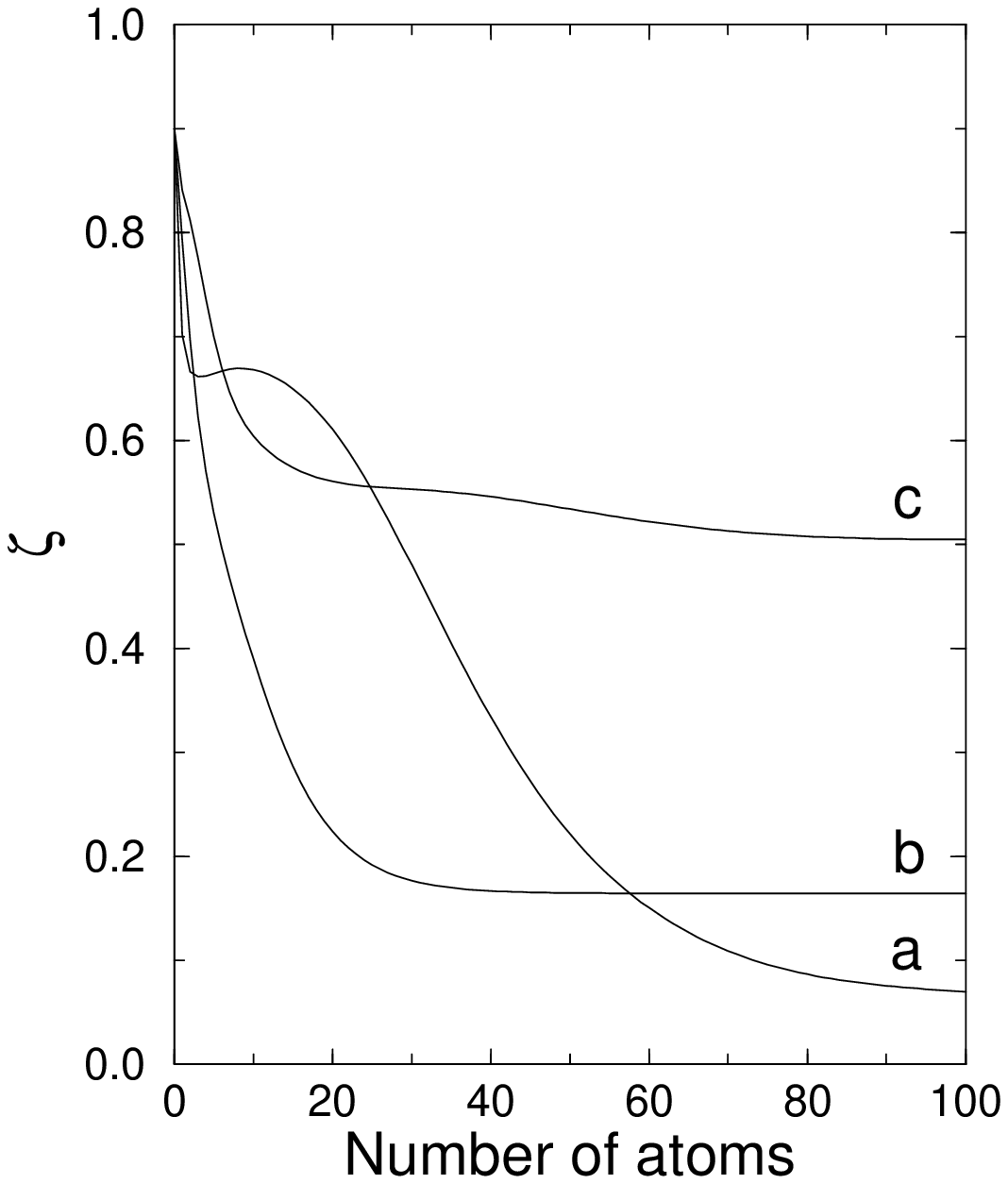}}
\end{center}
\vspace{0.5cm}
\caption{Field linear entropy $\zeta$ as a function of the number of atoms, for different classical external fields (a) $\varepsilon=1.0$, (b) $\varepsilon=2.0$ and (c) $\varepsilon=3.0$. The atom is initially 
in its excited state ($b=0$), the field in a thermal state having $\overline{n}=5$, and $\chi/\lambda = 
\Delta/\lambda = 1$.}
\label{fig2}
\end{figure}
have crossed the cavity, as shown in figure 2. In both cases there is an increase in the 
mean photon number of the cavity field (figure 3).

\subsection{Nonclassical properties}

The resulting field after the process of transfer of coherence occurs is not only an almost pure state,
but also has nonclassical features, for instance, statistical properties significantly different from the 
original thermal state. We are now going to calculate quantities which are relevant to characterize 
the generated state.

\subsubsection{Second order coherence function}

Quantum states of light may be classified according to their optical coherence properties. It is considered to be
a quantum regime if the one-mode second order coherence function $g^{(2)}(0)$ 
        \[
        g^{(2)}(0)=\frac{\langle \hat{a}^\dagger\hat{a}^\dagger\hat{a}\hat{a}\rangle}
        {\langle (\hat{a}^\dagger\hat{a})^2\rangle}
\]
\begin{figure}
\vspace{0.5cm}
\begin{center}
{\includegraphics [height=6truecm,width=4.7truecm]{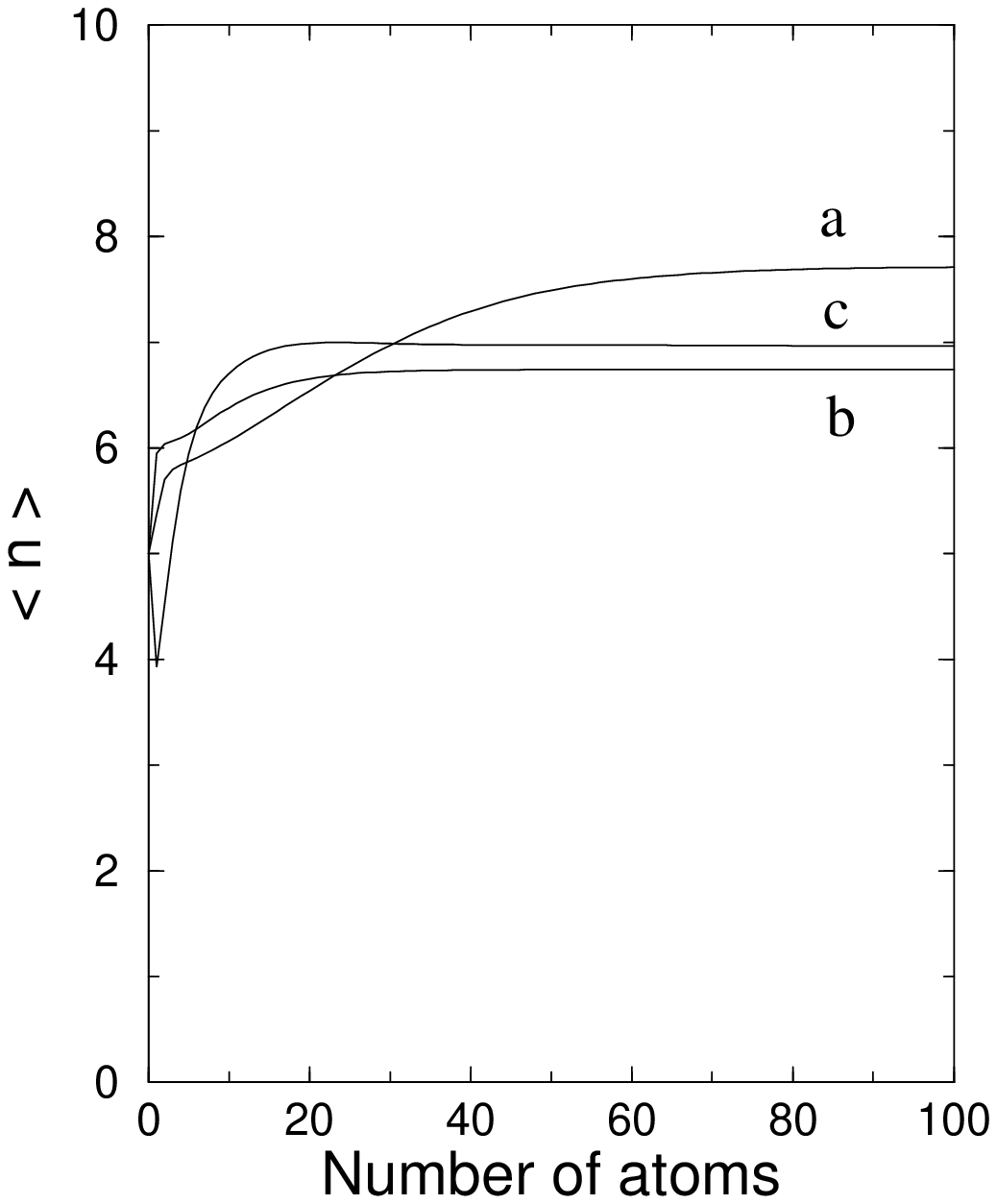}}
\end{center}
\vspace{0.5cm}
\caption{Mean photon number in the cavity as a function of the number of atoms. The same parameters
as in figure \ref{fig2}.}
\label{fig3}
\end{figure}

\begin{figure}
\vspace{0.5cm}
\begin{center}
{\includegraphics [height=6truecm,width=4.7truecm]{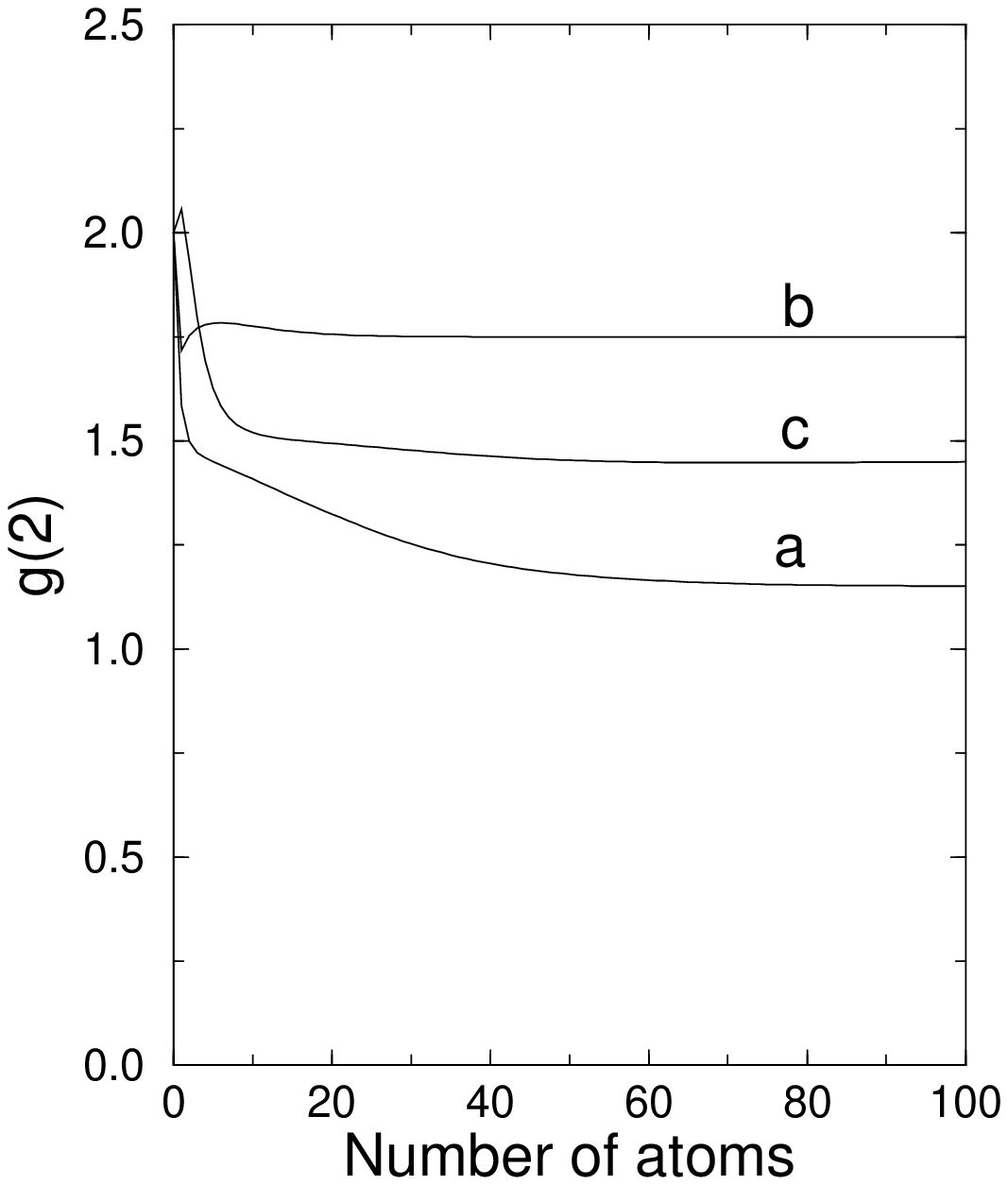}}
\end{center}
\vspace{0.5cm}
\caption{Second order correlation function of the cavity field as a function of the number of atoms. The same parameters as in figure \ref{fig2}.}
\label{fig4}
\end{figure}
is less than one. For a thermal field $g^{(2)}(0)=2$, and for a coherent field  $g^{(2)}(0)=1$. 
In our case we have verified that 
$g^{(2)}(0)$ decreases, although the exclusive quantum regime is never reached, as $g^{(2)}(0)$ is always greater than 
one. This may be seen in figure 4, where we have plotted $g^{(2)}(0)$ as a function of the number of atoms crossing the cavity.

\subsubsection{Representation in Fock and coherent state basis}

Projection of the field state in the Fock basis (photon number distribution, or 
$P_n=\langle n|\hat{\rho}|n\rangle$), may turn evident some nonclassical
properties. In this case we note a clear departure from the thermal (geometrical) distribution as
shown in figure 5, where it it shown the photon number distribution of the resulting field state after 
$N=100$ atoms have crossed the cavity. The photon number distribution of such a field, shown in figure 5, is 
very different from the distribution of a thermal state; we note strong oscillations in $P_n$, which is an 
evidence of nonclassical behaviour. This might be associated to the two-photon interaction employed in our scheme.
We remark that the photon number distribution gives us just partial 
information about the field. A more complete characterization may be obtained from quasiprobability functions in the coherent state basis, which are basically representations of the density operator $\hat{\rho}$ in terms of functions. 
A convenient quasiprobability is the $Q$-function, defined as \cite{wigner}
        \[
        Q(x,y)=\frac{1}{\pi}\langle \beta|\hat{\rho}|\beta\rangle,
\]
where $|\beta\rangle$ is a coherent state having amplitude $\beta=x+iy$. In figure 6 we have the $Q$-function
of the field after the passage of $N=100$ atoms. We note a double peaked structure, which resembles two
superposed deformed gaussians, which is displaced 
from the origin. As a matter of fact such a displacement may be attributed to the action of the classical field. This is
clearly seen if we change the sign of the classical field amplitude $\varepsilon$; for $\varepsilon=-1.0$, for instance,
the $Q$-function is displaced towards the opposite direction in phase space, as shown in figure 7, where
we have the contour plots of the $Q$-function for the case of $\varepsilon=-1.0$ compared to the case in which
$\varepsilon=1.0$.

\begin{figure}
\vspace{0.5cm}
\begin{center}
{\includegraphics [height=6truecm,width=4.7truecm]{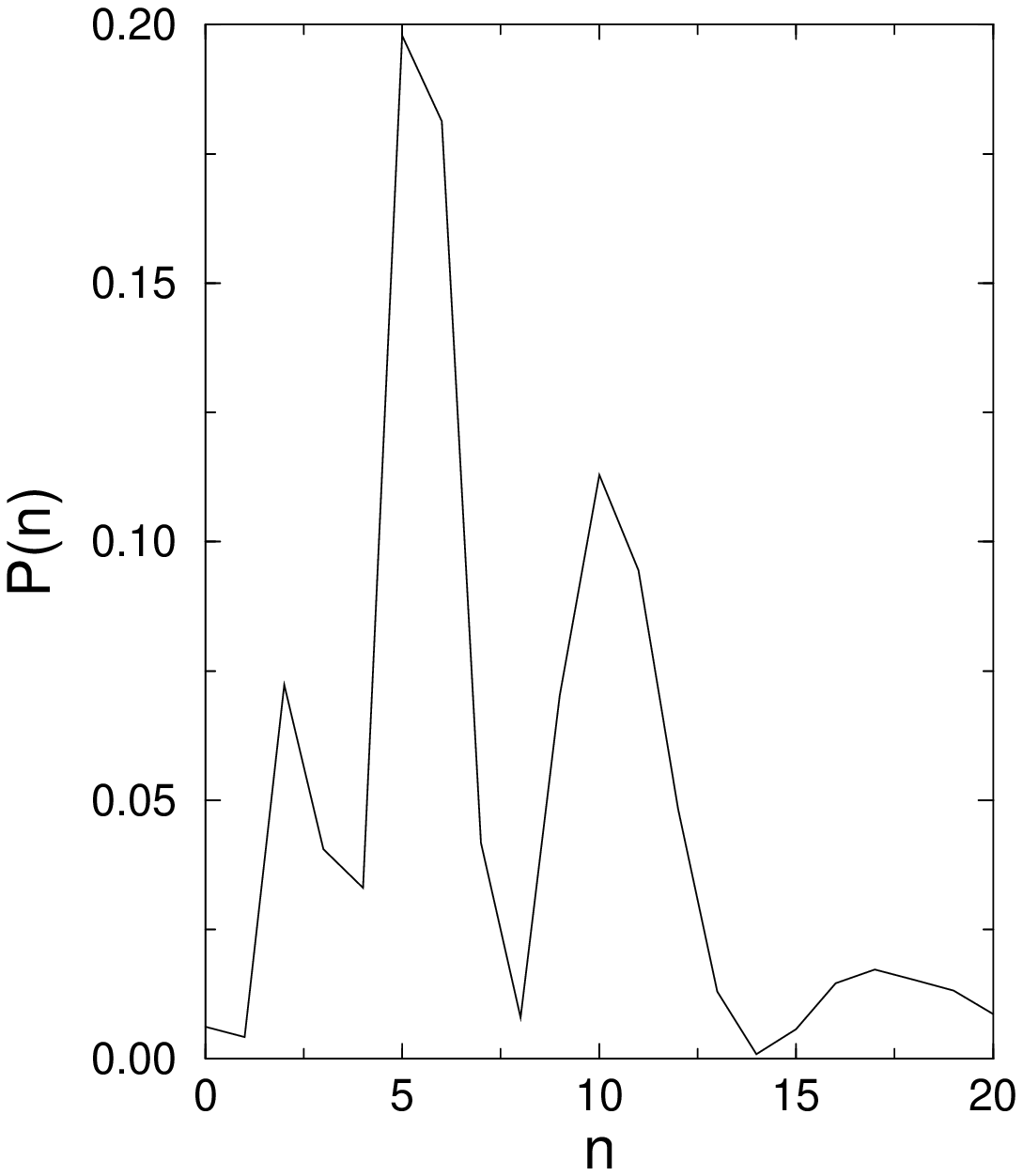}}
\end{center}
\vspace{0.5cm}
\caption{Photon number distribution of the cavity field after $N=100$ atoms have crossed the cavity and
with $\varepsilon=1.0$.}
\label{fig5}
\end{figure}

\begin{figure}
\begin{center}
{\includegraphics[height=7.5truecm,width=10.6truecm]{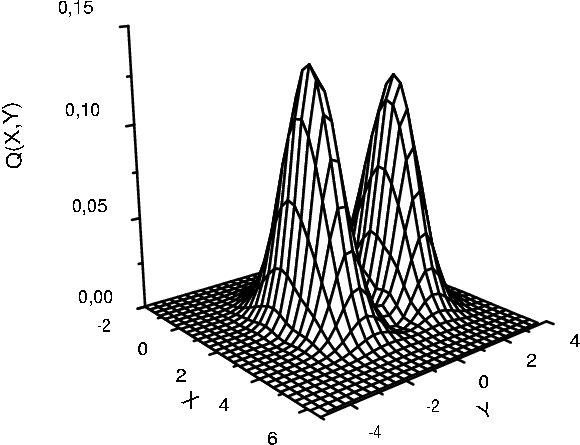}}
\end{center}
\caption{$Q$ function of the cavity field after $N=100$ atoms have crossed the cavity and
with $\varepsilon=1.0$.}
\label{fig6}
\end{figure} 
\begin{figure}
\begin{center}
{\includegraphics[height=6.0truecm,width=6.0truecm]{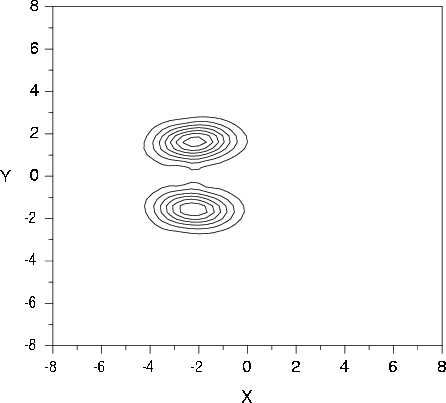}\hspace{1cm}
\includegraphics[height=6.0truecm,width=6.0truecm]{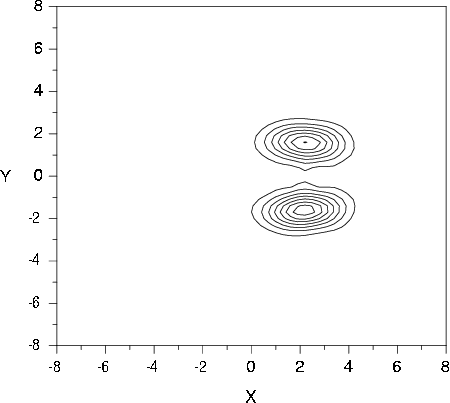}}
\end{center}
\caption{Contour plots of the $Q$ functions of the cavity field after $N=100$ atoms have crossed the cavity  with $\varepsilon=-1.0$ (in the left) and $\varepsilon=1.0$ (in the right).}
\label{fig7}
\end{figure}  

\section{Conclusion}

We have presented a scheme comprising a two-photon micromaser in which the transfer of coherence 
from atoms to a field in a statistical mixture (thermal state) is enhanced by coupling the atoms to a classical 
external field. We have shown that the continuous action of the external field has important effects on the scheme
of transfer of coherence based on a two-photon micromaser: it is required a simpler previous preparation 
of the atoms; the degree of purity of the field is considerably increased in a relatively short time, and  
there is no need of conditional measurements on the atoms leaving the cavity. 
The resulting field state is displaced from the origin in phase space, which may be interpreted as a direct 
consequence of the action of the external classical field. The field state has also nonclassical properties; 
we have verified strong oscillations in the photon number distribution of the field, a distinctive nonclassical 
feature, although there is no anti-bunching. A field with linear entropy as low as $\zeta\approx 0.07$ 
could be generated, in contrast to the case with no assistance of 
a classical field \cite{alvaro1}, which yields $\zeta\approx 0.53$. An important feature in our approach is that
the interaction times have been kept as short as possible, so that a time of $\approx 10^{-2}$ sec is long
enough to have around 100 atoms crossing the cavity. This is convenient if one wants to minimize the 
destructive effects of field dissipation, which normally leads to loss of coherence.
Here we wanted mainly to capture the main features of the model, e.g., the action of a classical external field, and we decided to investigate the dissipation effects elsewhere. It is though apparent that some kind of competition process between the atom-field transfer of coherence and dissipation-induced loss of coherence will be established if
cavity losses are taken into account. In summary, we have found that the process of transfer of coherence from one quantum subsystem (atoms) to another (electromagnetic field) may be substantially improved by applying an external  driving classical field. 

\section*{Acknowledgements} 

This work was partially supported by CNPq (Conselho Nacional para o 
Desenvolvimento Cient\'\i fico e Tecnol\'ogico), and FAPESP (Funda\c c\~ao 
de Amparo \`a Pesquisa do Estado de S\~ao Paulo), Brazil, and it is linked 
to the Optics and Photonics Research Center (FAPESP).



\begin{thebibliography}{0000}




\bibitem{haroche} M. Brune, S. Haroche, J.M. Raimond, L. Davidovich, N. Zagury, Phys. Rev. A 45 (1992) 5193. 

\bibitem{haroche1}  G. Nogues, A. Rauschenbeutel, A. Osnaghi, M. Brune, J.M. Raimond, S. Haroche, 
Nature 400 (1999) 239; G.R. Guth\"ohrlein, M. Keller, K. Hayasaka, W. Lange, H. Walther, 
Nature 414 (2001) 49.

\bibitem{rao} R.H. Xie and Q. Rao, Physica A 319 (2003) 233; R.H. Xie and Q. Rao, Physica A 315 (2002) 427; 
R.H. Xie and Q. Rao, Physica A 315 (2002) 386.

\bibitem{hector} C.A. Arancibia-Bulnes, H. Moya-Cessa, J.J. Sanchez-Mondragon, Phys. Rev. A 51 (1995) 5032.

\bibitem{alvaro1} A.F. Gomes, J.A. Roversi, A. Vidiella-Barranco, J. Mod. Opt. 46 (1999) 1421.

\bibitem{dutra} S.M. Dutra, P.L. Knight, H. Moya–Cessa, Phys. Rev. A 49 (1994) 1993. 

\bibitem{gao} F.L. Li, S.Y. Gao, Phys. Rev. A 62 (2000) 043809.

\bibitem{nha} H. Nha, Y.T. Chough, K. An, Phys. Rev. A 63 (2001) 010301.

\bibitem{wigner} D.F. Walls, G.J. Milburn, Quantum Optics, Springer-Verlag, Berlin 1994. 

\end{thebibliography}
\end{document}